# Universal Properties of Linear Magnetoresistance in Strongly Disordered Semiconductors

H.G. Johnson,[1] S.P. Bennett,[2] R. Barua,[3] L.H Lewis,[3] and D. Heiman[1]

[1]Department of Physics, Northeastern University, Boston, MA 02115
[2]Department of Mechanical Engineering, Northeastern University, Boston, MA 02115
[3]Department of Chemical Engineering, Northeastern University, Boston, MA 02115

*Abstract*

Linear magnetoresistance occurs in semiconductors as a consequence of strong electrical disorder and is characterized by nonsaturating magnetoresistance that is proportional to the applied magnetic field. By investigating a disordered MnAs-GaAs composite material, it is found that the magnitude of the linear magnetoresistance (LMR) is numerically equal to the carrier mobility over a wide range and is independent of carrier density. This behavior is complementary to the Hall effect that is independent of the mobility and dependent on the carrier density. Moreover, the LMR appears to be insensitive to the details of the disorder and points to a universal explanation of classical LMR that can be applied to other material systems.

## I. Introduction

Imperfections in materials, such as impurities and defects, alter their electronic structure in a variety of ways. These imperfections provide the backbone of many advanced technological applications, ranging from atomic substitutions in crystal lattices, to nanoscale pinning centers that lend functionality to superconductors and advanced magnetic materials. Although atomic substitution in electronic systems has been thoroughly investigated for more than half a century, research on strongly-disordered electronic systems is gaining increased attention. Of major interest is the role played by large-scale disorder on the magnetoelectronic properties of materials. Materials and devices with predictable magnetoresistance (MR) are of current interest for applications such as nonvolatile magnetic memories.[1] It is well known that in homogeneous semiconductors having limited electrical disorder, the magnetoresistance (MR) increases *quadratically* with increasing magnetic field, but eventually saturates at small fields. On the other hand, in compositionally inhomogeneous semiconductors having strong electrical disorder, the anomalously large MR increases *linearly* with increasing magnetic field, but does not saturate at high fields.[2,3,4,5] Materials with positive, nonsaturating linear magnetoresistance (LMR) comprise a distinct class of disordered semiconductors. For example, nonstoichiometric silver chalcogenide of the $Ag_{2+\delta}Se$ system exhibit a remarkably large LMR up to magnetic fields as high as H = 55 tesla.[4] The cause of this behavior is the presence of insulating or metallic particles embedded in the semiconductor matrix.

Several mechanisms have been proposed to characterize the phenomenon of LMR from geometrical,[6] classical,[7,8,9] quantum,[10,11] and effective medium[12,13] perspectives. The first model to be applied to LMR was a quantum description by Abrikosov, which examines a small bandgap material subjected to a magnetic field large enough to reach the quantum limit where all the electrons occupy the lowest Landau level.[10] This model also considered the existence of a linear energy dispersion relation similar to that found in single-layer graphite (graphene) at the Dirac point.[14] Alternatively, a classical model developed by Parish and Littlewood[7] examined the characteristic properties of LMR using a two-dimensional (2D)[7,8] resistor network that mimics inhomogeneous conducting media. This was subsequently expanded to a three-dimensional (3D) resistor network.[9] In these classical models the linear dependence of the magnetoresistance is linked in a simple manner to electrical disorder. A strong magnetic field forces a significant portion of the current to flow in a direction perpendicular to the applied voltage and thereby provides a linear-in-$H$ Hall resistance contribution to the effective magnetoresistance.

Here we describe the correlation of linear magnetoresistance to the fundamental electrical properties of a strongly disordered semiconductor system. The MR as a function of applied magnetic field is determined from the field-dependent transverse resistance $R(H)$ as

$$\mathrm{MR}(H) = [R(H) - R(H=0)] / R(H=0).$$

We find large LMR in a quasi-2D system of MnAs nanoparticles in a GaAs matrix. The magnitude of the LMR is found to be numerically equal to the magnitude of the macroscopic carrier mobility over a wide range of temperature and mobility. This equality, $\mathrm{MR}(T) = \mu(T)H$, is obeyed despite the fact that the carrier concentration changes by several orders of magnitude as a function of temperature. Since LMR is governed by carrier mobility and independent of carrier density, it is complementary to the classical Hall effect that is governed by the carrier concentration and is independent of carrier mobility. In addition, by examining different samples with a variety of disorder, the link between LMR and mobility does not appear to be affected by the specific details of the disorder, such as the size and concentration of the disordered regions. These results point to a universal description of MR in semiconductors with strong local disorder, offering insights for designing new MR systems.

## II. Experimental Setup and Electrical Properties

Transverse MR and Hall measurements were carried out on composite films[15] containing self-assembled MnAs metallic nanoparticles embedded in a GaAs matrix. Samples were fabricated from thin layers (20 - 50 nm) of homogeneous $Ga_{1-x}Mn_xAs$, x = 0.1, grown by low-temperature molecular beam epitaxy[16] (MBE) on GaAs substrates at a temperature of 250 ºC. MnAs nanoparticles formed in the films by self-assembly[17,18] when annealed at temperatures in the range 500 - 700 ºC for 30 minutes in the MBE chamber in an arsenic flux of vapor pressure of $10^{-6}$ torr. A variety of nanoparticle diameters and densities were synthesized with Mn content ranging from x = 0.06 to 0.10. Annealing temperatures in the range 570 - 670 ºC all led to a robust LMR response. The composite structures were examined by superconducting quantum interference device (SQUID) magnetometry, scanning electron microscopy (SEM), atomic force microscopy, and x-ray diffraction. The transverse MR was measured in magnetic fields applied perpendicular to both the current direction and the plane of the film. The inset of Fig. 1 is a SEM image showing MnAs nanoparticles with diameters ranging from 20 to 80 nm with an average of approximately 50 nm. We note that the metallic MnAs nanoparticles can be ferromagnetic or superparamagnetic, depending on the size of the particles, but the magnetism does not affect the MR for $T > 50$ K. However, it is the difference in the conductivity of the particles relative to the matrix that is essential in achieving LMR.

The electrical transport properties of a 20-nm-thick MnAs-GaAs composite film are shown in Fig. 1. Conduction in the film was found to be dominated by a high density of holes in the GaAs matrix. Figure 1a plots the density of the holes as a function of temperature, $p(T)$, obtained from Hall measurements. Near room temperature, the hole density is thermally activated from Mn acceptor sites of density $p_A = 4 \times 10^{19}$ cm$^{-3}$, with an activation energy of $E_A = 108$ meV.[19] For decreasing temperature the density decreases as the holes become trapped on the acceptors. Figure 1b plots the mobility of the holes as a function of temperature $\mu_h(T)$. Near room temperature, the hole mobility is $\mu_h = 100$ cm$^2$/Vs and increases for decreasing temperature due to reduced phonon scattering. Below $T \sim 50$ K the mobility decreases sharply and coincides with the rapid freeze out of the carriers.

### III. Magnetoresistance Results

The magnetoresistance of a composite film is plotted in Fig. 2 as a function of applied magnetic field $H$ for various temperatures. The transverse MR, which is positive and is an even function of the applied magnetic field, transitions from a quadratic field dependence at low fields to a linear dependence at higher fields. The high-field MR was found to be linear in $H$ at all temperatures 20 K $\leq T \leq$ 300 K. At the highest field, $H = 14$ T, the MR increased from 14 % at room temperature to nearly 1000% at $T = 25$ K.

The evolution of the MR with temperature is highlighted by plotting the temperature dependence of the *slope* of the linear portion of the MR response over the range $5 \leq H \leq 14$ T. Figure 3 shows a plot of the slope $d$MR/$dH$ as a function of inverse temperature, indicated by the solid circles. The straight line is a fit to the data and follows a power law dependence given by $d$MR/$dH \propto 1/T^{1.85}$. The seminal connection between the LMR and the carrier mobility is demonstrated by comparing the temperature dependencies of $d$MR/$dH$ and the hole mobility $\mu_h$, shown by solid squares. It is remarkable that the MR slope and the hole mobility have identical power-law temperature dependencies over the entire temperature range 50 K $\leq T \leq$ 300 K. Another striking similarity between the LMR and the carrier mobility is the saturation behavior at low temperatures, where both have a common maximum near $T \sim 40$ K. Below $T = 50$ K the mobility is no longer dominated by phonon scattering and the turnover of the data is commensurate with the rapid carrier freeze out. From Fig. 3 it is clear that the identical temperature dependencies of the LRM and the mobility unambiguously demonstrate that the magnetoresistance is *linearly proportional* to carrier mobility. Furthermore, by adjusting the scales to overlap the two sets of data, we obtain numerical equality between the magnitudes of the parameters, MR$(T) = \mu(T)H$, a remarkably universal result. It is also significant that this relation holds while the carrier density varies by several orders of magnitude, thus leading to the conclusion that the linear magnetoresistance is *independent* of carrier density.

Further evidence of the connection between the MR and carrier mobility is manifest in the crossover magnetic field, $H_{CR}(T)$, defined as the field marking the transition from quadratic to linear dependence in $H$. The crossover field was determined by fitting the MR($H$) data to a second-order polynominal in $H$.[20] The inset of Fig. 4 shows the region of the crossover for $T = 150$ K, where $H_{CR} = 1.15$ T. Using this method for MR experiments on Ag$_{2+\delta}$Se, the magnitude of the MR was found to be inversely proportional to the crossover field for a variety of sample thicknesses.[20] For our films we directly compare the crossover field with the mobility over a wide range of temperatures. Figure 4 shows a plot of $H_{CR}(T)$ and compares it to the temperature dependence of the inverse hole mobility. There is good agreement in the temperature dependencies of the crossover field and the inverse mobility, including the minima near $T = 40$ K. From these data it is apparent that for all temperatures the crossover field is linearly proportional to the inverse mobility, $H_{CR}(T) \propto 1/\mu(T)$, a further result that underscores the carrier mobility as the foundation of the MR behavior.

The experimental data for MnAs-GaAs composite films measured over a wide range clearly demonstrate that the magnetoresistance MR($T$) and crossover field $H_{CR}(T)$ are governed by the macroscopic carrier mobility $\mu(T)$. These attributes are comparable to predictions of a classical LMR model as articulated by Parrish and Littlewood.[8] In this model, the strongly inhomogeneous conductor is discretized into small spatial elements forming a random resistor network and analyzed numerically. A square lattice of four-terminal van der Pauw resistors was used, since two-terminal resistors are inadequate to describe a system containing a Hall effect component. For an infinite network of nonidentical resistors the MR was found to be linear in $H$ for a magnetic field directed perpendicular to the lattice. This model predicts that the MR is dominated either by the average mobility, $\langle\mu\rangle$, or by the width of the mobility distribution, $\Delta\mu$. For a wide mobility distribution, $\Delta\mu/\langle\mu\rangle > 1$, both MR and $H_{CR}$ are expected to be a function of the width of the mobility variation $\Delta\mu$. This behavior was assigned to $Ag_{2+\delta}Se$, where the average mobility vanished, $\langle\mu\rangle = 0$, as a result of having equal proportions of electron and hole regions.[9] On the other hand, in the regime of percolating conductivity between silver grains in $Ag_{2+\delta}Se$ a narrow mobility width, $\Delta\mu/\langle\mu\rangle < 1$, was found.[20] For a narrow mobility distribution, $\Delta\mu/\langle\mu\rangle < 1$, both MR and $H_{CR}$ are expected to be functions of the average carrier mobility. In the present MnAs-GaAs composite system the observed dependence of LMR on the average mobility indicates a narrow mobility distribution, $\Delta\mu/\langle\mu\rangle < 1$.

Finally, we note a connection between the mean free path of the carriers and length scale of the disorder. At higher temperatures, $T > 50$ K, the carrier mean free path is dominated by phonon scattering and $\ell_{mfp} \sim 10\text{-}10^2$ nm. Here, the mean free path is smaller than the average MnAs nanoparticle separation in the system, $\langle d \rangle \sim 10^2$ nm, and results in the close overlap of the LMR and mobility seen in Fig. 3 above T = 50 K. On the other hand, at low temperatures, $T < 50$ K, the mean free path becomes larger than the average particle separation, $\ell_{mfp} > \langle d \rangle$, and the LMR is observed to deviate from the mobility. At this point the classical model also breaks down.[8] Another essential feature of the present study is the weak connection between the LMR and the electrical disorder. In the regime where the mobility is small and dominated by phonon scattering it was found that the magnitude of the LMR was essentially identical for several samples possessing different particle sizes and densities. This result indicates that the details of the disorder are not important for generating LMR as long as the mean free path of the carriers is smaller than the length scale of the system inhomogeneities.

### IV. Conclusions

The results presented here on magnetotransport in MnAs-GaAs composite films reveal that nonsaturating linear magnetoresistance is governed strictly by the average macroscopic carrier mobility. Furthermore, the LRM is complementary to the Hall effect – whereas the LMR is proportional to mobility and independent of carrier density, the Hall voltage is independent of mobility and inversely proportional to the carrier density. For applications, a two-terminal LMR device can be advantageous over a four-terminal Hall effect device. An important feature of LMR is the insensitivity of the magnetotransport behavior to the degree of the random disorder. These observations provide convincing support for a universal model of a strongly inhomogeneous semiconductor possessing random disorder in the conductivity. This universality renders it applicable to a wide range of materials.

## Acknowledgments

This work supported by grant DMR-097007 from the National Science Foundation. We thank W. Fowle for expertise with SEM imaging and B. Chaprut for expertise with scanning probe imaging.

[1] A. Fert, Rev. Mod. Phys. **80**, 1517 (2008).
[2] R. Xu, A. Hussmann, T.F. Rosenbaum, M.-L. Saboungi, J.E. Enderbya, and P.B. Littlewood, Nature (London) **390**, 57 (1997).
[3] S.L. Bud'ko, P.C. Canfield, C.H. Mielke, and A.H. Lacerda, Phys. Rev. B **57**, 13624 (1998).
[4] A. Husmann, J.B. Betts, G.S. Boebinger, A. Migliori, T.F. Rosenbaum, and M.-L. Saboungi, Nature (London) **417**, 421 (2002).
[5] Jingshi Hu and T.F. Rosenbaum, Nature Matl. **7**, 697 (2008).
[6] W.R. Branford, A. Husmann, S.A. Solin, S.K. Clowes, T. Zhang, Y.V. Bugoslavsky, and L.F. Cohen, Appl. Phys. Lett. **86**, 202116-1 (2005).
[7] M.M. Parish and P.B. Littlewood, Nature (London) **426**, 162 (2003).
[8] M.M. Parish and P.B. Littlewood, Phys. Rev. B **72**, 094417 (2005).
[9] J. Hu, M.M. Parish, and T.F Rosenbaum, Phys. Rev. B **75**, 214203 (2007).
[10] A.A. Abrikosov, Phys. Rev. B **58**, 2788 (1998).
[11] A.A. Abrikosov, Europhys. Lett. **49**, 789 (2000).
[12] V. Guttal and D. Stroud, Phys. Rev. B **71**, 201304(R) (2005).
[13] V. Guttal and D. Stroud, Phys. Rev. B **73**, 085202 (2006).
[14] A.A. Abrikosov, Phys. Rev. B **60**, 4231 (1999).
[15] R.S. DiPietro, H.G. Johnson, S.P. Bennett, T.J. Nummy, L.H. Lewis, and D. Heiman *Appl. Phys. Lett.* **96**, 222506 (2010).
[16] H. Ohno, A. Shen, F. Matsukura, A. Oiwa, A. Endo, S. Katsumoto, and Y. Iye, Appl. Phys. Lett. **69**, 363 (1996).
[17] J. De Boeck, R. Oesterholt, A. Van Esch, H. Bender, C. Bruynseraede, C. Van Hoof, and G. Borghs, Appl. Phys. Lett. **68**, 2744 (1996).
[18] M. Moreno, A. Trampert, B. Jenichen, L. Daweritz, and K.H. Ploog, J. Appl. Phys. **92**, 4672 (2002).
[19] B. Lakshmi, G. Favrot, and D. Heiman, Proc. SPIE **5359**, 290 (2004).
[20] M. von Kreutzbruck, G. Lembke, B. Mogwitz, C. Korte, and J. Janek, Phys. Rev. B **79**, 035204 (2009).

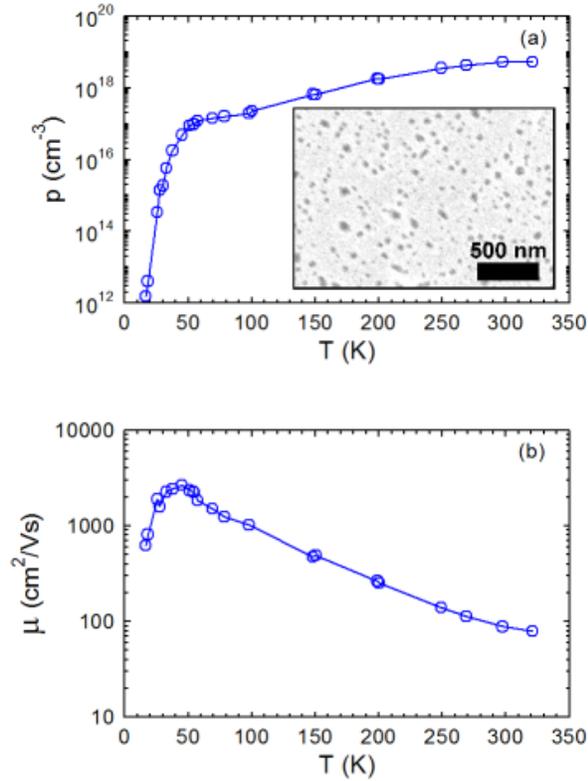
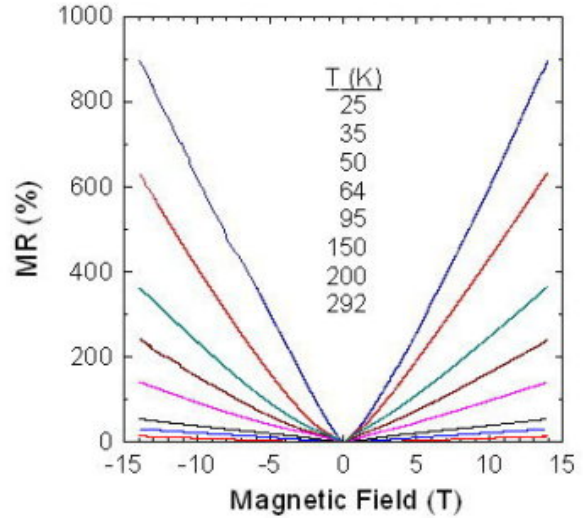

Fig. 1. Temperature-dependent electrical transport properties of MnAs-GaAs self-assembled composite films. (a-inset) SEM micrograph of a self-assembled composite film depicting MnAs nanoparticles of diameter 20 nm - 80 nm (50 nm average diameter) dispersed in the GaAs matrix. (a) Graph of density of mobile holes versus temperature, $p(T)$, obtained from Hall effect measurements, highlighting carrier freeze out at low temperature. (b) Graph of hole Hall mobility versus temperature, $\mu_h(T)$, showing a slowly increasing carrier mobility with decreasing temperature, from the room-temperature value of $\mu_h = 100$ cm$^2$/V-s.

Fig. 2. Transverse magnetoresistance (MR) of MnAs-GaAs self-assembled composite film versus applied magnetic field ($H$) at various temperatures ($T$). The magnetic field is applied perpendicular to the plane of the film in which the current was flowing. For increasing field, the variation of the MR is proportional to $H^2$ and quickly becomes linear in field at higher fields. For increasing temperature the MR increases from a value of MR = 14 % at room temperature to 900% at $T$ = 25 K.

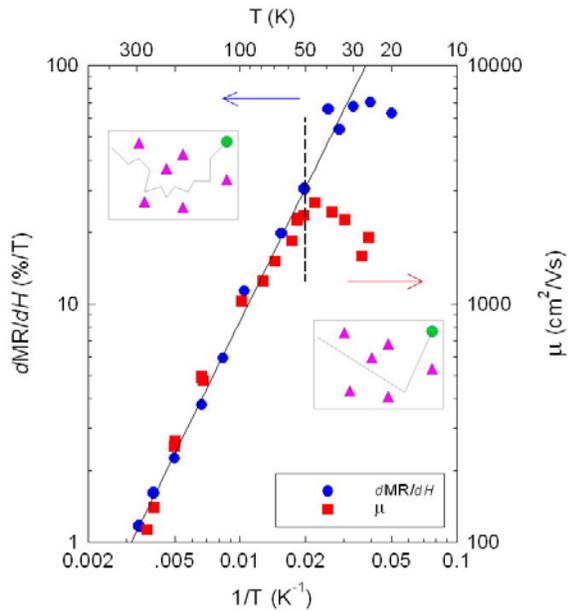
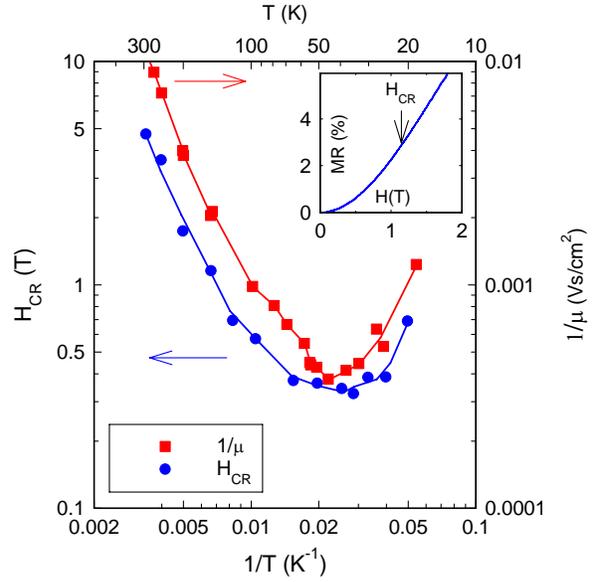

Fig. 3. Linear magnetoresistance and hole mobility of MnAs-GaAs self-assembled composite film plotted as a function of inverse temperature, $1/T$. The magnitude of the linear magnetoresistance, expressed as the derivative of the MR at high fields, $d\text{MR}/dH$, is shown by the blue solid circles. The hole mobility, $\mu_h(T)$, obtained from Hall effect measurements, is shown by the red solid squares and has been scaled to overlay the magnetoresistance values. Both of the quantities have identical temperature dependencies, $d\text{MR}/dH(T) = \mu_h(T)$, and the solid line through the data is a power law $\propto T^{-1.85}$. The dashed line marks the temperature where the carrier mean free path is equal to the average nanoparticles spacing, illustrated in the insets.

Fig. 4. Crossover field, $H_{CR}$, and inverse hole mobility, $1/\mu_h$, of a MnAs-GaAs self-assembled composite film versus inverse temperature, $1/T$. $H_{CR}$ is defined where the magnetoresistance crosses over from a quadratic $H^2$-dependence at low fields to a linear $H$-dependence at high fields, shown in the inset for T = 150 K. $H_{CR}(T)$ is shown by the blue solid circles. The inverse hole mobility obtained from Hall effect measurements is shown by the red solid squares and has been scaled to lie close to the crossover field values. The quantities have closely matched temperature dependencies, $H_{CR}(T) \propto 1/\mu_h(T)$.